\documentclass[prl, preprint, superscriptaddress]{revtex4-1}

\usepackage{amsmath}    
\usepackage{graphicx}   
\usepackage{verbatim}   
\usepackage{color}      
\usepackage{hyperref}   
\usepackage{SIunits}    
\usepackage{longtable}

\usepackage[textsize=scriptsize]{todonotes}

\begin{document}

\title{Tetrahedral colloidal clusters from random parking of
  bidisperse spheres}

\author{Nicholas B. Schade}
\affiliation{Department of Physics, Harvard University, Cambridge,
  Massachusetts 02138, USA}

\author{Miranda C. Holmes-Cerfon}
\affiliation{School of Engineering and Applied Sciences, Harvard
  University, Cambridge, Massachusetts 02138, USA}

\author{Elizabeth R. Chen}
\affiliation{Department of Mathematics, University of Michigan, Ann
  Arbor, Michigan 48109, USA}

\author{Dina Aronzon}
\affiliation{School of Engineering and Applied Sciences, Harvard
  University, Cambridge, Massachusetts 02138, USA}

\author{Jesse W. Collins}
\affiliation{School of Engineering and Applied Sciences, Harvard
  University, Cambridge, Massachusetts 02138, USA}

\author{Jonathan A. Fan}
\affiliation{School of Engineering and Applied Sciences, Harvard
  University, Cambridge, Massachusetts 02138, USA}

\author{Federico Capasso}
\affiliation{School of Engineering and Applied Sciences, Harvard
  University, Cambridge, Massachusetts 02138, USA}

\author{Vinothan N. Manoharan}
\affiliation{School of Engineering and Applied Sciences, Harvard
  University, Cambridge, Massachusetts 02138, USA}
\affiliation{Department of Physics, Harvard University, Cambridge,
  Massachusetts 02138, USA}

\date{\today}

\begin{abstract}
  Using experiments and simulations, we investigate the clusters that
  form when colloidal spheres stick irreversibly to -- or ``park'' on
  -- smaller spheres. We use either oppositely charged particles or
  particles labeled with complementary DNA sequences, and we vary the
  ratio $\alpha$ of large to small sphere radii.  Once bound, the
  large spheres cannot rearrange, and thus the clusters do not form
  dense or symmetric packings.  Nevertheless, this stochastic
  aggregation process yields a remarkably narrow distribution of
  clusters with nearly 90\% tetrahedra at $\alpha=2.45$.  The high
  yield of tetrahedra, which reaches 100\% in simulations at
  $\alpha=2.41$, arises not simply because of packing constraints, but
  also because of the existence of a long-time lower bound that we
  call the ``minimum parking'' number. We derive this lower bound from
  solutions to the classic mathematical problem of spherical covering,
  and we show that there is a critical size ratio
  $\alpha_c=(1+\sqrt{2})\approx 2.41$, close to the observed point of
  maximum yield, where the lower bound equals the upper bound set by
  packing constraints. The emergence of a critical value in a random
  aggregation process offers a robust method to assemble uniform
  clusters for a variety of applications, including metamaterials.
\end{abstract}

\maketitle

Understanding the geometry of clusters formed from small particles is
a fundamental problem in condensed matter physics, with implications
for phenomena ranging from nucleation \cite{Kelton03} to self-assembly
\cite{Manoharan03}.  Colloidal particles are a useful experimental
system for studying cluster geometry and its relation to phase
behavior \cite{Gasser01} for several reasons: they are large enough to
be directly observed using optical microscopy; their assembly can be
understood in terms of geometry \cite{Arkus09, Hoy10}; and they can be
driven to cluster by a variety of controllable interactions, including
capillary forces \cite{Manoharan03}, depletion \cite{Meng10},
fluctuation-induced forces \cite{Yunker11}, or DNA-mediated attraction
\cite{Soto02}.  Colloidal clusters are also useful materials in their
own right.  They can be used, for example, as building blocks for
isotropic optical metamaterials known as metafluids \cite{Urzhumov07,
  Alu09, Fan10}.  Tetrahedral clusters are of particular interest for
metafluids since the tetrahedron is the simplest cluster with
isotropic dipolar symmetry \cite{Urzhumov07}. An unsolved challenge
for this application is to determine the interactions and conditions
that enable assembly of bulk quantities of highly symmetric, uniform
clusters such as tetrahedra.

\begin{figure}[t]
\begin{center}
\includegraphics[width=8.6cm]{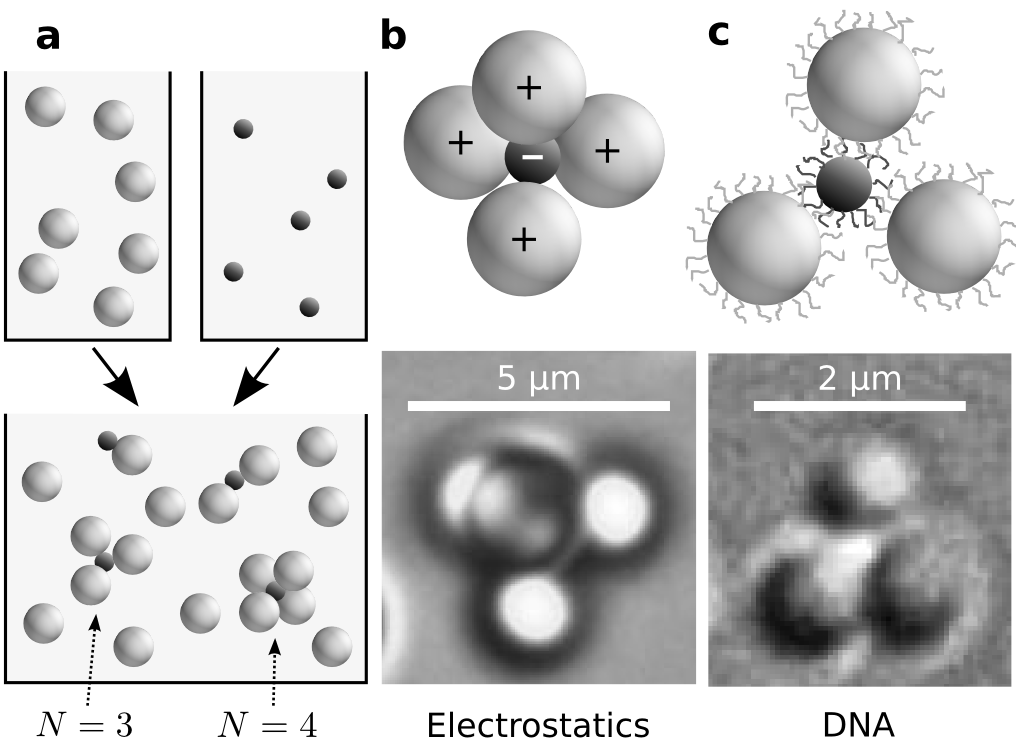}
\caption{\label{fig:experimentalSystem6}(a) Two colloidal sphere
  species are mixed together to form clusters. (b) Oppositely charged
  polystyrene spheres cluster due to electrostatic attraction. Optical
  micrograph shows a tetramer ($N=4$). (c) Polystyrene spheres labeled
  with complementary DNA strands (not to scale) cluster due to DNA
  hybridization. Optical micrograph shows a trimer ($N=3$); the small,
  central sphere is fluorescent.  
}
\end{center}
\end{figure}

With this motivation in mind, we study experimentally the geometry and
size distribution of binary clusters formed when small colloidal
spheres are mixed with an excess of large spheres that stick
\emph{irreversibly and randomly} to their surfaces
(Figure~\ref{fig:experimentalSystem6}a).  An obvious way to control
the cluster geometry in such binary systems is to vary the size
ratio. One might expect that at certain ratios the particles could
arrange into dense clusters or ``spherical packings'' -- arrangements
of spheres around a central sphere that maximize surface density
\cite{Melnyk77, Conway93, Phillips12}.  Such packings have long been
used in modeling the microstructure of dense, disordered atomic
systems \cite{Egami97, Miracle04}.  But unlike atoms, colloidal
particles can stick irreversibly, such that two particles bound to a
third show no motion relative to one another.  This type of binding
occurs frequently in strongly interacting, monodisperse colloidal
suspensions, which consequently form fractal aggregates instead of
dense glasses \cite{Weitz84, Lin89}.  Similarly, in the binary systems
we study, the irreversible and stochastic process of sticking
precludes the formation of dense or symmetric packings.  The large
spheres \emph{park}, rather than pack, on the surfaces of the small
spheres.

Surprisingly, this random and non-equilibrium process can produce
clusters of uniform size.  Our experiments show that at a size ratio
$\alpha = R_{\mathrm{big}}/R_{\mathrm{small}} = 2.45$, where
$R_{\mathrm{big}}$ and $R_{\mathrm{small}}$ are the sphere radii,
nearly all of the clusters contain four large spheres stuck to a
smaller sphere (Table~\ref{table:histograms}). In these experiments we
use a 100:1 stoichiometric ratio of the two sphere species,
statistically ensuring  that each cluster contains only one small
sphere surrounded by two or more larger spheres.  After waiting
several days for the average cluster size to saturate, we measure the
distribution of $N$, the number of large spheres bound to each small
sphere \footnote[1]{See Supplementary Information at end of manuscript for additional
  details.}.  We do not count single large spheres, nonspecifically
aggregated clusters of large spheres, or clusters with multiple small
spheres.  While there are many isolated large spheres due to the high
stoichiometric ratio, the latter two types of cluster are rare.

\begin{table}[t]
\begin{tabular}{c|r r r r}
Size ratio $\alpha$ & 1.94 & \color{red} 2.45 & 3.06 & 4.29 \\ [0.5ex]
\hline
$N=6$ &  6.3 & \color{red}  0.0 &  0.0 &  0.0 \\
$N=5$ & 39.2 & \color{red}  0.8 &  0.0 &  0.0 \\
$N=4$ & 54.4 & \color{red} 90.2 & 18.6 &  0.7 \\
$N=3$ &  0.0 & \color{red}  6.6 & 69.9 & 35.9 \\
$N=2$ &  0.0 & \color{red}  0.8 & 10.9 & 51.0 \\
$N=1$ &  0.0 & \color{red}  0.8 &  0.6 & 11.1 \\
$N=0$ &  0.0 & \color{red}  0.8 &  0.0 &  1.3 \\ [1ex]
\end{tabular}
  \caption{\label{table:histograms} Experimentally observed cluster size
    distributions for charged colloids. Percentages of total are
    listed. The distribution for $\alpha=2.45$ (red) is sharply peaked
    at $N=4$.} 
\end{table}

The $N=4$ tetramers that we observe are not dense packings or, in
general, symmetric arrangements. As can be seen from the images in
Figure~\ref{fig:experimentalSystem6}, there is space between the large
particles, and the resulting tetrahedra are irregular.  Moreover, the
ratio $\alpha=2.45$ is well below the value $\alpha = 4.44$ found by
Miracle \textit{et al.} \cite{Miracle03} for efficient tetrahedral
packing in binary atomic clusters.  In fact, at $\alpha=4.29$, closer
to this bound, we see much smaller clusters and few tetrahedra.  The
sparsity of large spheres in the clusters is a result of the
irreversible, non-equilibrium, random binding: once the big particles
stick to the smaller ones, we do not see them detach or move relative
to one another.  We expected such a stochastic process to lead to a
much broader distribution of clusters.  At other values of $\alpha$ it
does (Table~\ref{table:histograms}), but at $\alpha=2.45$ we obtain
90\% tetramers.

The high yield of tetramers occurs in two experimental systems with
different types of interactions.  In both systems the interactions are
specific, strong, and short-ranged, and the particles do not rearrange
once bound.  In the first system the clustering is driven by
electrostatic interactions.  We mix large, positively-charged
particles with small, negatively-charged particles, as shown in
Figure~\ref{fig:experimentalSystem6}b.  To adjust $\alpha$, we use
several different particle sizes \footnote[1]{}.  We add salt to
reduce the Debye length to approximately \unit{3}{\nano \meter}, small
enough to ensure that the interaction range does not significantly
influence the effective particle size.  In the second system the
clustering is driven by hybridization of grafted DNA strands
\footnote[1]{}.  As shown in Figure~\ref{fig:experimentalSystem6}c, we
mix small and large spheres labeled with complementary DNA
oligonucleotides \cite{Dreyfus09}.  We work well below the DNA melting
temperature so that the attractive interaction is many times the
thermal energy \cite{Dreyfus10}.

To better understand why the distribution is sharply-peaked at $N=4$
for $\alpha = 2.45$, we use simulations and analytical techniques that
account for the irreversibility of the aggregation process.  Our
simulations use a ``random parking'' algorithm \cite{Mansfield96,
  Rosen86, Wouterse05, Talbot00} to model the formation of clusters.
The algorithm involves attaching large spheres to randomly selected
positions on the surface of a small sphere, subject to a no-overlap
constraint \footnote[1]{}.  We do not model the finite range of the
interactions, which in both experimental systems is small compared to
the particle size, or the diffusion of the particles prior to binding.
In accord with experimental observations, the particles are not
allowed to rearrange once bound.  We repeat the process numerically to
obtain distributions of cluster sizes as a function of a single
parameter, $\alpha$.

\begin{figure}[t]
\begin{center}
\includegraphics[width=8.6cm]{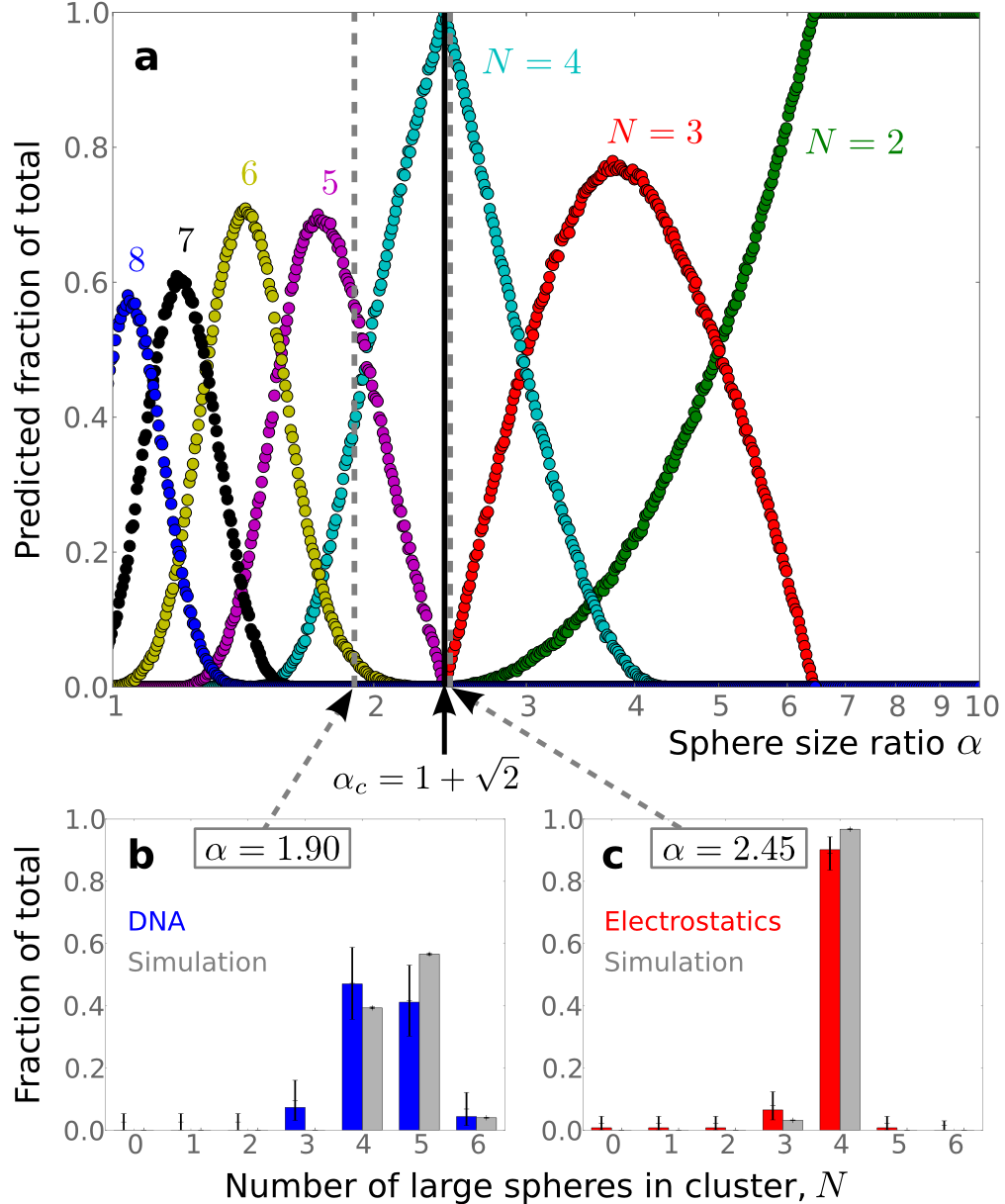}
\caption{\label{fig:histograms4} (a) Yield curves, as determined by
  simulations, for $N$-particle clusters, $2 \le N \le 8$, where the
  critical size ratio $\alpha_c$ is marked with a black line.
  Below are histograms for (b) DNA-labeled particles (blue) at $\alpha
  = 1.90$ and (c) charged particles (red) at $\alpha = 2.45$, as
  observed in experiments and as predicted from simulations (gray).
  Error bars are 95\% confidence intervals (Wilson score interval
  method).}
\end{center}
\end{figure}

The simulations find a 100\% yield of tetramers at the size ratio
$\alpha \approx 2.41$.  As in the experiments, the large particles in
these tetramers are not densely packed, and the clusters are therefore
distorted tetrahedra.  We also find that while the yield of any
particular cluster can be maximized by varying $\alpha$
(Figure~\ref{fig:histograms4}a), the yield approaches 100\% only for
dimers ($N=2$) and tetramers ($N=4$).  Interestingly, the yield curve
for tetramers has a cusp at its peak, showing that the size ratio
$\alpha_c$ at the maximum is a mathematical critical point.

The simulated distributions agree well with those found experimentally
(Figure~\ref{fig:histograms4}b,c) for both electrostatic and
DNA-mediated interactions.  For instance, at $\alpha=2.45$ with
electrostatic interactions, we find a sharply-peaked distribution
consisting almost entirely of tetramers.  This value of $\alpha$ is
close to but not precisely at the critical value, so a small yield of
trimers is predicted and observed experimentally.  In contrast, at
$\alpha=1.90$ we find a mixture of mostly $N=4$ and $N=5$ clusters in
both the DNA system and simulations.  Some discrepancy arises between
the simulated and experimental histograms because the yield curves in
Figure~\ref{fig:histograms4}a are steep; a slight error in the
effective size ratio can shift the cluster distribution. Nevertheless,
the random sphere parking model successfully reproduces both the large
yield of tetrahedra near $\alpha_c$ and the details of the measured
histograms at various other $\alpha$.

That we can reproduce the same phenomenon in two different
experimental systems and in a one-parameter model suggests that the
critical size ratio $\alpha_c$ has a universal, geometrical origin.
Intuitively one might expect that it is related to packing constraints
on the large spheres.  Other theoretical studies of random sphere
parking \cite{Rosen86, Mansfield96} have calculated the maximum number
of large spheres $N_{\max}$ that can fit around a small sphere at a
given $\alpha$.  However, this bound cannot by itself explain why the
yield of tetramers can reach 100\% while that of other clusters, such
as trimers or hexamers, cannot.  At a given $\alpha$, it tells us only
why no clusters larger than $N_{\max}(\alpha)$ can form, but it says
nothing about the probability of forming smaller clusters with
different arrangements.

Therefore we also examine a different bound, one not previously
discussed in the context of random sphere parking: the ``minimum
parking'' curve $N_{\min}(\alpha)$.  $N_{\min}$ is the smallest number
of hard spheres that can be positioned on a smaller sphere such that
another sphere cannot fit.  To understand this bound, consider a
simple, one-dimensional analogy to car parking on a busy city street,
where if a space opens up that is large enough to fit a car, it is
filled.  The minimum parking number occurs when all drivers have been
equally inconsiderate, leaving spaces between their parked cars that
are all slightly too small for another car to fit. This lower bound is
meaningful only at long times, when all available parking spaces have
been filled.  The long-time limit holds also for our experiments and
simulations, which we carry out until the average cluster size has
saturated.

Whereas the upper bound $N_{\max}(\alpha)$ is straightforwardly
related to solutions of the well-known spherical packing problem
\cite{Conway93, Sloane11}, the calculation of the lower bound
$N_{\min}(\alpha)$ requires a different approach.  In our clusters,
the distance between the centers of any two big spheres must be at
least $2R_{\mathrm{big}}$. Consider then a sphere of radius
$(R_{\mathrm{small}}+R_{\mathrm{big}})$ that circumscribes the centers
of the parked spheres.  If this sphere is completely covered with $N$
circles of radius $2R_{\mathrm{big}}$, it will be impossible to add an
$(N+1)^{\mathrm{th}}$ large sphere.  We are led naturally to the
\emph{spherical covering} problem, a problem with a rich history in
mathematics.  Like spherical packings, spherical coverings are
solutions to an extremum problem: they are arrangements of $N$ points
on a sphere that minimize the largest distance between any location on
the sphere surface and the closest point \cite{Conway93}.  But unlike
spherical packings, spherical coverings need not correspond to
arrangements of non-overlapping spheres.  We therefore solve for the
minimum parking curve by examining the solutions to the spherical
covering problem \cite{Sloane11} at each $N$ and manually verifying
that they correspond to non-overlapping configurations \footnote[1]{}.

\begin{figure}[t]
\begin{center}
\includegraphics[width=8.6cm]{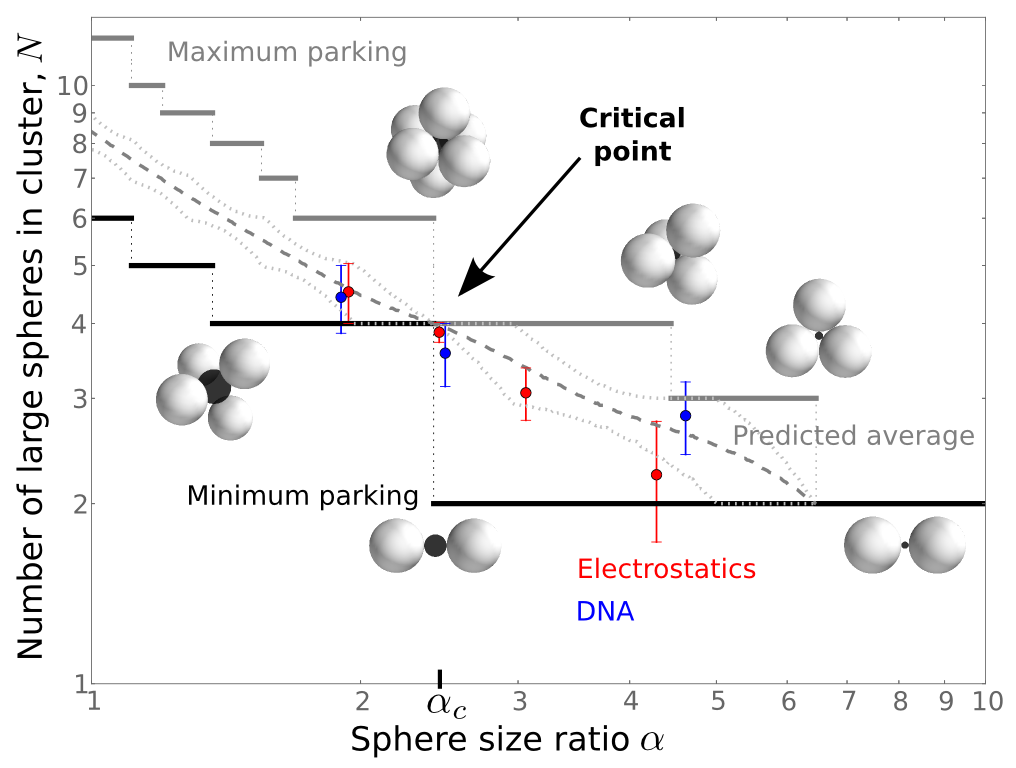}
\caption{\label{fig:expAndTheory4} $N_{\max}$ (solid gray) and
  $N_{\min}$ (black) as functions of $\alpha$. Cluster images show
  sphere configurations at discontinuities of these curves.  Average
  cluster sizes from simulations (dashed gray) and experiments (blue
  and red data points) are shown.  We characterize the statistical
  dispersion in each distribution by the average absolute deviation
  from the median, indicated by dotted light gray lines for
  simulations and vertical bars for experiments.}
\end{center}
\end{figure}

Our analytical results for the bounds reveal why $\alpha_c$ is a
special point: it is the only non-trivial point where the calculated
maximum and minimum parking curves come together
(Figure~\ref{fig:expAndTheory4}).  Analytically we find the location
of the critical value to be $\alpha_c=(1+\sqrt{2})\approx 2.41$, very
close to the values where the experimental distributions are
peaked. At $\alpha$ slightly larger than this value, the minimum
parking configuration corresponds to two spheres placed at opposite
poles ($N_{\min}=2$), and the maximum $N$ is obtained by first parking
three large spheres next to one another, so that there is room for one
more sphere to park ($N_{\max}=4$).  At $\alpha$ slightly smaller than
$\alpha_c$, the big spheres can park along orthogonal axes about the
small sphere to make an octahedron ($N_{\max}=6$).  The minimum $N$ is
obtained by placing four spheres as far from each other as possible,
so as to make the addition of a fifth impossible ($N_{\min}=4$).  Thus
as we increase $\alpha$ through $\alpha_c$, $N_{\max}$ goes from $6$
to $4$ and $N_{\min}$ from $4$ to $2$, and the two curves become
infinitesimally close.

The parking process is therefore geometrically constrained to yield
clusters with exactly $N=4$ particles in the limit
$\alpha\rightarrow\alpha_c$. A simple geometric argument sheds some
light on this result. At $\alpha_c$ there is always room for four
large spheres to park. Parking more spheres requires that at least
three park precisely along a great circle of the smaller particle, but
the probability of this happening randomly is zero.  Thus irreversible
binary aggregation, a stochastic process, has a deterministic feature
at the critical size ratio: although the space between the large
spheres can vary, all clusters must be tetramers.  Our numerical
approach confirms that the statistical dispersion in the cluster size
distribution vanishes at $\alpha_c$, as shown in
Figure~\ref{fig:expAndTheory4}.

The experimental and simulated distributions differ slightly due to
two effects.  First, the measured sizes tend to be smaller than the
simulated ones because a few parking spaces remain unfilled even at
long times.  This effect is more pronounced for larger spheres, which
diffuse more slowly and encounter the small spheres less frequently.
The systems most affected are the electrostatic ones at $\alpha =
3.06$ and $4.29$.  Second, the experimental size ratios can vary by
5\% due to polydispersity. Both of these factors increase the width in
the experimental distributions and diminish the achievable yield of
tetramers near $\alpha_c$.  Nevertheless, the experimental data
indicate that near $\alpha_c$ a tetramer yield of at least 90\% is
possible, and, although the model does not account for the finite
range of the interactions or the diffusivity of the particles, it is
useful for predicting cluster size distributions in two very different
colloidal systems.

These results have both fundamental and practical consequences.  On
the fundamental side, the particle size ratio could affect the jamming
threshold in bulk packings of bidisperse spheres.  Previous
simulations of these systems have shown that the distribution of
coordination numbers also depends on the size ratio \cite{He99} and
may be modeled using random parking \cite{Wouterse05}.  This contrasts
with dense atomic systems like metallic glasses \cite{Egami97,
  Miracle04} in which the atoms have some freedom to rearrange
locally.  In these systems packing constraints may explain structure
and coordination better than parking arguments.

On the practical side, this random aggregation process is a simple way
to mass produce tetrahedral clusters in theoretically 100\%
yield. Although the tetrahedra we produce are irregular in that the
distance between the large spheres can vary, it may well be possible
to form large quantities of symmetric tetrahedra simply by shrinking
the small spheres after the tetramers have formed \cite{Lyon04}.
Furthermore, although the yield will approach 100\% only for dimers
and tetramers, the yield of any $N$-particle cluster can be maximized
by choosing the appropriate size ratio.  For instance, the yield of
octahedral clusters, also promising candidates for building
metamaterials \cite{Alu09}, may surpass 70\% at $\alpha = 1.42$.
 
The size ratio in binary colloidal systems thus emerges as a valuable
control parameter for directed self-assembly. Moreover, because it
does not require precise control over the interactions, random parking
offers a robust and simple way to make colloidal clusters that are
more monodisperse than those prepared through other methods
\cite{Manoharan03}.

We thank W. Benjamin Rogers, Rodrigo Guerra and Michael P. Brenner for
helpful discussions.  This work was funded by the National Science
Foundation NIRT program (grant no. ECCS-0709323) and the Harvard MRSEC
(grant no. DMR-0820484). NBS acknowledges support from the Department
of Energy Office of Science Graduate Fellowship Program, administered
by ORISE-ORAU under contract no. DE-AC05-06OR23100.

\bibliography{schade-arxiv}

\pagebreak

\section{Electrostatic interaction experiments}
Charged colloidal polystyrene spheres were purchased from Invitrogen
as ``IDC surfactant-free latex'' in batches as listed in
Table~\ref{table:stocks}.

\begin{table}[h]
\begin{tabular}{c c c r}
Mean diameter & Surface functionality & Fluorescent? & Surface charge \\ [0.5ex]
\hline
\unit{0.49}{\micro \meter} & carboxylate-modified latex (CML) & yes & \unit{-262}{\micro \coulomb / \centi \meter ^2} \\
\unit{0.95}{\micro \meter} & amidine & no & \unit{+23.7}{\micro \coulomb / \centi \meter ^2} \\
\unit{1.2}{\micro \meter} & amidine & no & \unit{+18.2}{\micro \coulomb / \centi \meter ^2} \\
\unit{1.5}{\micro \meter} & aldehyde-amidine & no & \unit{+18.2}{\micro \coulomb / \centi \meter ^2} \\
\unit{2.1}{\micro \meter} & amidine & no & \unit{+30.2}{\micro \coulomb / \centi \meter ^2} \\ [0.5ex]
\end{tabular}
  \caption{\label{table:stocks} Charged particles used in
    electrostatic system experiments. Values for surface charge from
    data sheets provided by manufacturers.}
\end{table}

A \unit{100}{\micro \liter} sample of each colloid was diluted to 1\%
weight by volume.  This was then vortexed for a few seconds and
bath-sonicated for 10 seconds.  We cleaned the particles by
centrifuging and redispersing them several times in deionized (DI)
water, using the following wash procedure.

\begin{enumerate}
\item Colloids were centrifuged for 5 minutes at 6600g.
\item Supernatant was removed and \unit{190}{\micro \liter} of DI
  water were added to each sample.
\item Samples were vortexed for 5 seconds each and then bath-sonicated
  for 10 seconds.
\end{enumerate}

We performed six wash cycles.  After the last centrifugation, the
supernatant was replaced with \unit{40}{\micro \liter} of DI water,
rather than \unit{190}{\micro \liter} as before.  Then
\unit{50}{\micro \liter} of 20 mM NaCl were added to each sample to
achieve an overall salt concentration of 10 mM.  This screens the
repulsion between like particles before mixing.

We prepared mixtures of the positively and negatively charged
particles such that each mixture contained one batch of positively
charged particles at 1\% w/v.  In each mixture, the number ratio of
the large (positively charged) to small (negatively charged) spheres
was $100 : 1$.  The salt concentration in each mixture was 10 mM
NaCl.  Each mixture consisted of large and small particles with a
different size ratio $\alpha$, as listed in
Table~\ref{table:mixtures}.

\begin{table}[h]
\begin{tabular}{c l l}
$\alpha$ & Large particles & Small particles \\ [0.5ex]
\hline
$1.94$ & \unit{0.95}{\micro \meter} amidine (+) & \unit{0.49}{\micro
  \meter} CML (-) \\
$2.45$ & \unit{1.2}{\micro \meter} amidine (+) & \unit{0.49}{\micro
  \meter} CML (-) \\
$3.06$ & \unit{1.5}{\micro \meter} aldehyde-amidine (+) &
\unit{0.49}{\micro \meter} CML (-) \\
$4.29$ & \unit{2.1}{\micro \meter} amidine (+) & \unit{0.49}{\micro
  \meter} CML (-) \\ [0.5ex]
\end{tabular}
  \caption{\label{table:mixtures} Size ratios and components of
    binary mixtures of charged colloids.}
\end{table}

Each mixture was stored in a micro-centrifuge tube and vortexed at
3,000 RPM, bath-sonicated for 20 seconds, and then mounted on a
Glas-Col Rugged Rotator to tumble slowly at 4$^\circ$C.  Each mixture
tumbled for at least three days before observation to reduce the
effects of sedimentation on local particle concentrations throughout
the mixture. To make it easier to identify and characterize one
cluster at a time, we diluted samples of the mixtures to 0.1\% w/v
just prior to observing the distribution of cluster sizes.

\subsection{Electrostatics control experiment}

In the experiments outlined above, each mixture contained particles
with surface charges of opposite sign.  In a separate control
experiment, we mixed particles of two different sizes but with surface
charges of the same sign.  Both components in our control mixture were
carboxylate-modified latex (CML) colloids with a size ratio $\alpha =
2.24$, as listed in Table~\ref{table:elec_control}.

\begin{table}[h]
\begin{tabular}{c c c r}
Mean diameter & Surface functionality & Fluorescent? & Surface charge \\ [0.5ex]
\hline
\unit{0.49}{\micro \meter} & CML & yes & \unit{-262}{\micro \coulomb / \centi \meter ^2} \\
\unit{1.1}{\micro \meter} & CML & no & \unit{-31.5}{\micro \coulomb / \centi \meter ^2} \\ [0.5ex]
\end{tabular}
  \caption{\label{table:elec_control} Colloids used in electrostatic
    system control experiment.}
\end{table}

\begin{figure}[h]
\begin{center}
\includegraphics[width=8.6cm]{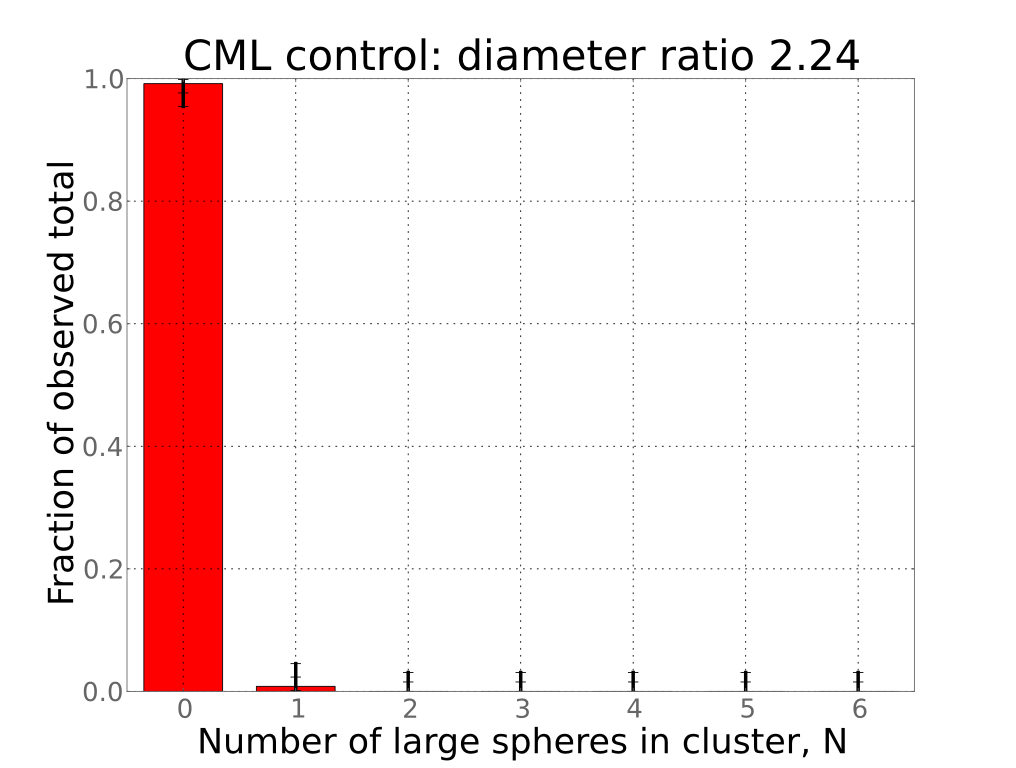}
\caption{\label{fig:electrostaticControl1} Cluster size distribution
  in a mixture of \unit{1.1}{\micro \meter} non-fluorescent CML
  particles and \unit{0.49}{\micro \meter} fluorescent CML particles
  in a $100 : 1$ number ratio, showing that particles with surface
  charge of the same sign rarely form clusters.}
\end{center}
\end{figure}

These colloids were washed using the procedure outlined above and then
mixed in a $100 : 1$ number ratio.  After tumbling at 4$^\circ$C for
several days, the cluster size distribution was measured.  As shown in
Figure~\ref{fig:electrostaticControl1}, fewer than 1\% of the small
particles bind to large particles when they have surface charges of
the same sign.  Nonspecific aggregation is therefore rare in the
charged colloidal systems.

\subsection{Experiments without salt}

In another set of experiments, mixtures of charged colloidal particles
were prepared as described above, but without salt.  The number ratio
of positively charged spheres to negatively charged spheres was again
$100 : 1$, and several size ratios $\alpha$ were investigated, as
listed in Table~\ref{table:no_salt_mixtures}.  As in the experiments
with 10 mM NaCl, we tumbled the mixtures for several days before
measuring the distribution of clusters.

\begin{table}[h]
\begin{tabular}{c l l}
$\alpha$ & Large particles & Small particles \\ [0.5ex]
\hline
$1.09$ & \unit{1.2}{\micro \meter} amidine (+) & \unit{1.1}{\micro
  \meter} CML (-) \\
$1.36$ & \unit{1.5}{\micro \meter} aldehyde-amidine (+) &
\unit{1.1}{\micro \meter} CML (-) \\
$1.94$ & \unit{0.95}{\micro \meter} amidine (+) & \unit{0.49}{\micro
  \meter} CML (-) \\
$2.45$ & \unit{1.2}{\micro \meter} amidine (+) & \unit{0.49}{\micro
  \meter} CML (-) \\
$3.06$ & \unit{1.5}{\micro \meter} aldehyde-amidine (+) &
\unit{0.49}{\micro \meter} CML (-) \\
$4.29$ & \unit{2.1}{\micro \meter} amidine (+) & \unit{0.49}{\micro
  \meter} CML (-) \\ [0.5ex]
\end{tabular}
  \caption{\label{table:no_salt_mixtures} Size ratios and components of
    binary mixtures without salt.} 
\end{table}

\begin{figure}[h]
\begin{center}
\includegraphics[width=8.6cm]{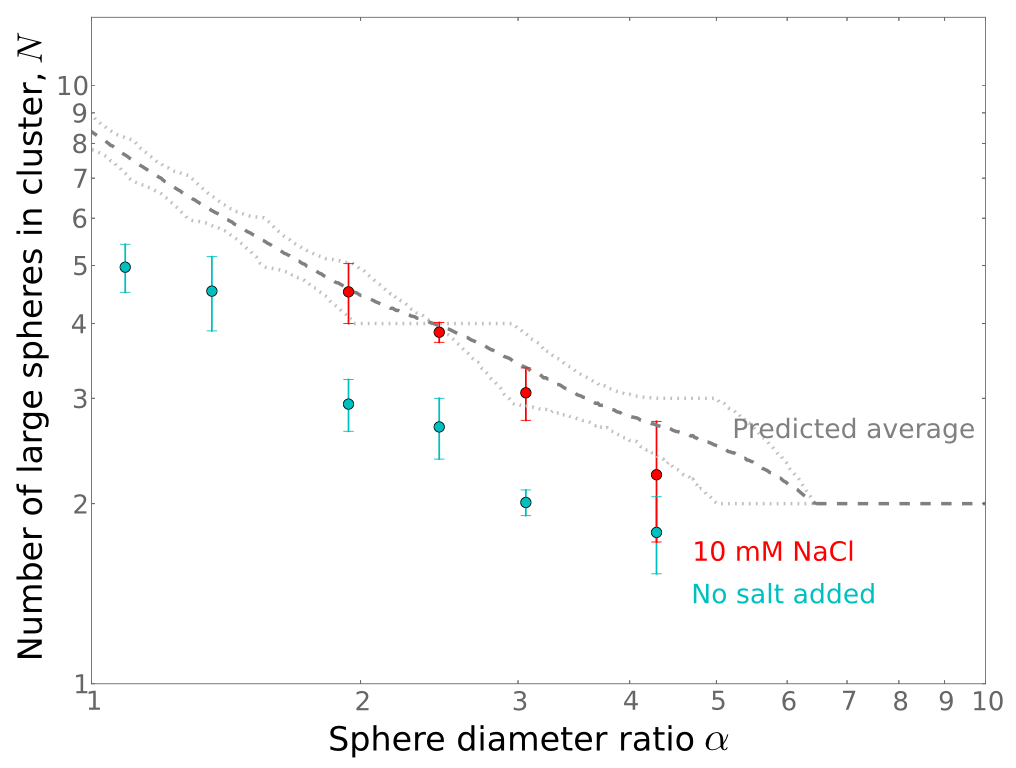}
\caption{\label{fig:electrostaticSaltComp1} Average cluster sizes from
  simulations (dashed dark gray) and electrostatic experiments  with
  (red data points) and without (cyan) salt.  Widths of the cluster
  size distributions are indicated by dotted light gray lines for
  simulations and vertical error bars for experiments.}
\end{center}
\end{figure}

Figure~\ref{fig:electrostaticSaltComp1} shows that the average cluster
sizes in these mixtures were smaller than the average sizes predicted
from simulation and those observed in mixtures containing 10 mM NaCl.
For instance, at $\alpha=2.45$ with 10 mM NaCl the average cluster size
is $N = 3.9$, but when there is no salt in the system the average
cluster size is $N = 2.7$.  For the four size ratios for
which there is data both without salt and with 10 mM NaCl, we found
that clusters are 20\% to 35\% smaller when no salt is added.

These experiments show that electrostatic repulsion affects the
cluster assembly.  This observation is consistent with other recent
experiments \cite{Wang10} showing that the cluster size distribution
in binary mixtures depends on ionic strength when the salt
concentration is less than 10 mM.  In our mixtures with 10 mM NaCl,
the Debye length is approximately \unit{3}{\nano \meter}, very small
compared to the particle sizes, so the large spheres do not interact
with each other except at small distances.  The random parking model
should be more appropriate for systems like these where the
interaction range is much smaller than the particle size.  This is
because the random parking model assumes no interactions between the
particles except for a hard-core repulsion and irreversible binding on
contact between spheres of two different types.

\pagebreak

\section{DNA-colloid experiments}

In another set of experiments, we mixed small and large spheres
labeled with complementary 65-base ssDNA oligonucleotides purchased
from Integrated DNA Technologies:
\begin{itemize}
\item Sequence A: 5'-biotin-51xT-TGTTGTTAGGTTTA-3'
\item Sequence B: 5'-biotin-51xT-TAAACCTAACAACA-3'
\end{itemize}

The oligonucleotides terminate with a biotin group, which allows us to
graft them to streptavidin-coated polystyrene particles using a
protocol from Dreyfus \textit{et al.} \cite{Dreyfus09}.  The
streptavidin-coated polystyrene particles are purchased from Bangs
Laboratories, Inc. with the following diameters:
\begin{itemize}
\item \unit{0.21}{\micro \meter}, fluorescent, to be coated with sequence B
\item \unit{0.39}{\micro \meter}, fluorescent, to be coated with sequence B
\item \unit{0.51}{\micro \meter}, fluorescent, to be coated with sequence B
\item \unit{0.97}{\micro \meter}, non-fluorescent, to be coated with sequence A
\end{itemize}

DNA strands were dissolved in water in \unit{20}{\micro}M
concentration.  \unit{20}{\micro \liter} of this DNA solution were
mixed with \unit{10}{\micro \liter} of 1\% w/w streptavidin-coated
polystyrene particles with \unit{120}{\micro \liter} of phosphate
buffer in a 1.7 mL propylene micro-centrifuge tube.  Each 50 mL batch
of buffer contained 0.0128 g KH$_2$PO$_4$, 0.0707 g K$_2$HPO$_4$,
0.1467 g NaCl, and 0.250 g F108 surfactant in 50 mL of deionized
water.  The buffer was filtered through a \unit{0.2}{\micro \meter}
membrane before use.  A separate batch containing no salt was also
prepared so that the salt concentration could be adjusted by combining
the two.  Mixtures were vortexed at 3,000 RPM for 5 seconds and
bath-sonicated for 10 seconds. They were incubated at room temperature
for 30 minutes to allow the ssDNA to graft to the surface of the
particles.  Then we washed the colloids using the following procedure:

\begin{enumerate}
\item Colloids were centrifuged for 3 minutes at 12,000g.
\item Supernatant was removed and \unit{100}{\micro \liter} of 50 mM
  NaCl buffer were added to each sample.
\item Samples were vortexed for 5 seconds each at 3,000 RPM and then
  bath-sonicated for 10 seconds.
\end{enumerate}

We washed each colloid three times to remove excess DNA from the
system, and then incubated each at 55$^\circ$C for 30 minutes.  We
then washed three more times, incubated at 55$^\circ$C for
another 30 minutes, and washed three times again.  At this point
the salt concentration in the buffer was adjusted to 20 mM.  The A-
and B-labeled particles were mixed in a $100 : 1$ (large : small)
number ratio such that the larger particles were at a volume fraction
of about 0.1\%.  Three separate mixtures were prepared, as listed in
Table~\ref{table:DNAmixtures}.

\begin{table}[h]
\begin{tabular}{c l l}
$\alpha$ & Large particles & Small particles \\ [0.5ex]
\hline
$1.90$ & \unit{0.97}{\micro \meter}, sequence A & \unit{0.51}{\micro
  \meter}, sequence B \\
$2.49$ & \unit{0.97}{\micro \meter}, sequence A & \unit{0.39}{\micro
  \meter}, sequence B \\
$4.62$ & \unit{0.97}{\micro \meter}, sequence A & \unit{0.21}{\micro
  \meter}, sequence B \\ [0.5ex]
\end{tabular}
  \caption{\label{table:DNAmixtures} Size ratios and components of
    mixtures with DNA-driven interactions.} 
\end{table}

After the mixtures had been prepared, we followed the same procedure
that we used for the charged colloid system.

\subsection{DNA-colloid control experiment}

In the experiments outlined above, each mixture contained particles
labeled with complementary DNA strands.  In a separate control
experiment, we mixed particles of two different sizes but labeled with
identical ssDNA.  

\begin{figure}[h]
\begin{center}
\includegraphics[width=8.6cm]{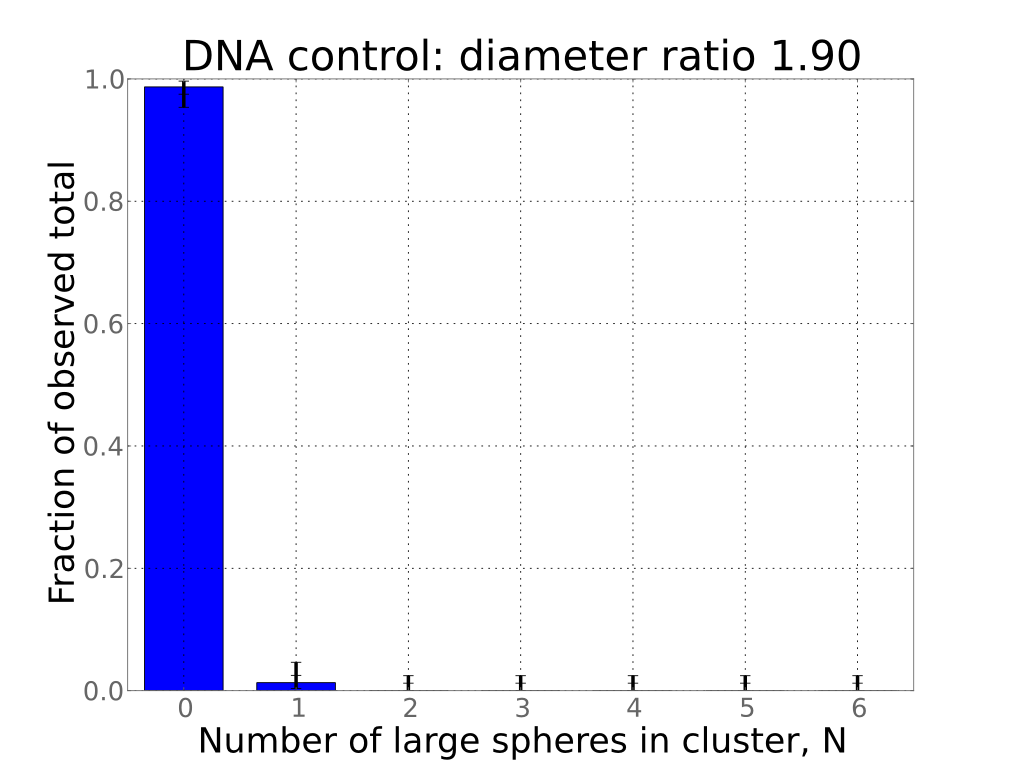}
\caption{\label{fig:DNAcolloidControl1} Cluster size distribution in a mixture of \unit{0.97}{\micro \meter} non-fluorescent particles and \unit{0.51}{\micro \meter} fluorescent particles in a $100 : 1$ number ratio, coated with the same non-self-complementary DNA sequence (A).}
\end{center}
\end{figure}

Both components in our control mixture were streptavidin-coated
polystyrene colloids labeled with sequence A.  We used
\unit{0.97}{\micro \meter} (non-fluorescent) and \unit{0.51}{\micro
  \meter} (fluorescent) particles, yielding a size ratio $\alpha =
1.90$.  The particles were functionalized with DNA using the same
procedure described above and then mixed in a $100 : 1$ number ratio.
After tumbling at $4^\circ$C for several days, the cluster size
distribution was measured. Figure~\ref{fig:DNAcolloidControl1} shows
that fewer than 2\% of the small spheres bind to large spheres when
they are coated with the same DNA sequence.  The low amount of
nonspecific aggregation is expected, since we designed sequence A to
have a negligible amount of self-hybridization even at $0^\circ$C.

\section{Measurement of cluster size distribution}

To measure the distribution of cluster sizes of a particular mixture,
we placed a \unit{5}{\micro \liter}, 0.1\% w/v sample between two
cover slips and then sealed it at the edges with UV-curable epoxy
(Norland Optical Adhesive 61).  We observed each sample under
differential interference contrast with a 100X oil-immersion objective
on a Nikon Eclipse TE2000-E inverted microscope.  We used fluorescence
to identify the small spheres.

The distribution was obtained by counting the number of clusters at
each $N$.  $N=0$ is a small sphere without any large spheres adsorbed,
$N=1$ a small sphere attached to a single large sphere, etc.  We also
counted clusters containing multiple small spheres and clusters formed
through non-specific aggregation.  The number of such clusters is
small compared to the total number of clusters counted.  Thus, we do
not include these in the histograms. 

Our counting procedure is designed to avoid errors from double
counting.  We begin by counting clusters in the field of view (FOV) on
the microscope, scanning through $z$ to find clusters that may
initially be out of focus.  The Brownian motion of each cluster
eventually brings all particles into view.  We can therefore determine
the number of large and small spheres in each cluster through direct
observation.  Once we have recorded all clusters in the FOV, we
translate the FOV to another part of the sample located more than one
FOV away.  We repeat this process to build up the histogram, rastering
the FOV over the sample.  We count more than 60 clusters at each size
ratio and more than 120 clusters at size ratios greater than $2$.  The
entire histogram for a given mixture is recorded in one session on the
microscope.

\section{Algorithm for simulating random sphere parking}

We simulate the cluster distribution using an algorithm based on Monte
Carlo trials.  These consist of two stages:

\paragraph{Coarse stage}  
We repeatedly try to insert a disc of radius $r = R_\mathrm{big} /
(R_\mathrm{big}+R_\mathrm{small})$ on the surface of the unit
sphere. The center of the disc is randomly and uniformly distributed
on the sphere.  If the disc overlaps any discs that are already
``parked'', we reject it; otherwise we add it to the list of parked
discs. After a fixed number of consecutive rejections, (typically
$N_{\text{coarse}} = 10^4$), we switch to the fine stage.

\paragraph{Fine stage}
We compute the remaining regions of possible insertion, and then try
to insert an additional disc.  We choose the center of the disc
randomly and uniformly from these regions.

To find the regions of possible insertion, we first compute the arcs
that form their boundaries: Given a central disc of radius $r$,
neighboring disc centers must lie on or outside a concentric circle of
radius $2r$. We erase arcs that lie in the interior of the concentric
circles of radius $2r$ about each neighboring disc. If the central
circle contains any unerased arcs, these are added to our roster of
arcs.

We do this for each parked disc center, and then stitch together the
remaining arcs to form the regions to which another disc center could
be added.  To find the area of each region, we inscribe the region in
a circumcircle and use a Monte-Carlo integration method to determine
the ratio of the area of the region to the area of its circumcircle.

The algorithm terminates when the list of remaining arcs in the fine
stage is empty, at which point the final number of inserted discs is
recorded.

\section{Calculating bounds on cluster size distribution}

The upper and lower bounds $N_{\max}$ and $N_{\min}$ on the cluster size
distribution can be calculated from \textit{spherical codes}
\cite{Sloane11} corresponding to known solutions to the problems of
spherical packings ($N_{\max}$) and spherical coverings ($N_{\min}$).
Spherical packings are arrangements of $N$ points on a unit sphere
that maximize the smallest distance between any two of them
\cite{Conway93, Sloane11}.  For a given $\alpha$ we determine
$N_{\max}(\alpha)$ by looking up the spherical packing \cite{Sloane11}
with the largest $N$ for which the minimal distance between points is
at least $2\alpha / (1 + \alpha)$. Our calculation of
$N_{\min}(\alpha)$ is more involved, as explained below.

\subsection{Connection between spherical covering and minimum parking}

Here we verify that the best known coverings are also optimal
solutions to the minimum parking problem. For a given configuration of
$n$ ``parked'' points on a unit sphere, let the \emph{covering radius}
be the maximum distance between any point on the sphere to the nearest
parked point. Let the \emph{packing radius} be (half) the minimum of
the pairwise distances between the parked points. If the parked points
represent centers of circles with some radius $r$, it is impossible to
add another circle without overlapping a parked one if and only if the
covering radius is less than $2r$. The circles that are already parked
do not overlap provided that the packing radius is greater than
$r$. Therefore, if we are given an optimal solution to the covering
problem, that is, one that minimizes the covering radius for a given
$n$, it will also be the optimal minimal parking configuration if the
covering radius is less than 2 times the packing radius.

We manually verify that the best known coverings satisfy this
constraint for $n=4, \ldots, 130$ by calculating the packing radius
for each optimal known covering, obtained from ref. \cite{Sloane11}. Table
\ref{tbl:coverings} shows the covering radius, packing radius, and
difference (in degrees). The difference is always positive, confirming
our statement.

\begin{longtable}{r c c p{3cm}}
     n &      2 * packing radius &     covering radius &   2*(packing radius) - (covering radius)\\ [0.5ex]
     \hline \\
     4 &      109.47 &       70.53 &       38.94\\ 
     5 &       90.00 &       63.43 &       26.57\\ 
     6 &       90.00 &       54.74 &       35.26\\ 
     7 &       72.00 &       51.03 &       20.97\\ 
     8 &       61.76 &       48.14 &       13.62\\ 
     9 &       68.97 &       45.88 &       23.09\\ 
    10 &       65.53 &       42.31 &       23.22\\ 
    11 &       50.65 &       41.43 &        9.22\\ 
    12 &       63.43 &       37.38 &       26.06\\ 
    13 &       46.23 &       37.07 &        9.16\\ 
    14 &       52.58 &       34.94 &       17.64\\ 
    15 &       45.67 &       34.04 &       11.63\\ 
    16 &       50.48 &       32.90 &       17.58\\ 
    17 &       41.63 &       32.09 &        9.54\\ 
    18 &       45.53 &       31.01 &       14.51\\ 
    19 &       40.73 &       30.37 &       10.36\\ 
    20 &       40.01 &       29.62 &       10.39\\ 
    21 &       39.45 &       28.82 &       10.62\\ 
    22 &       40.70 &       27.81 &       12.89\\ 
    23 &       38.99 &       27.48 &       11.51\\ 
    24 &       36.67 &       26.81 &        9.86\\ 
    25 &       36.75 &       26.33 &       10.42\\ 
    26 &       35.12 &       25.84 &        9.27\\ 
    27 &       38.06 &       25.25 &       12.81\\ 
    28 &       35.87 &       24.66 &       11.21\\ 
    29 &       33.86 &       24.37 &        9.50\\ 
    30 &       31.18 &       23.88 &        7.30\\ 
    31 &       29.94 &       23.61 &        6.33\\ 
    32 &       37.38 &       22.69 &       14.69\\ 
    33 &       26.61 &       22.59 &        4.02\\ 
    34 &       30.82 &       22.33 &        8.49\\ 
    35 &       27.98 &       22.07 &        5.90\\ 
    36 &       28.78 &       21.70 &        7.08\\ 
    37 &       31.22 &       21.31 &        9.91\\ 
    38 &       30.31 &       21.07 &        9.24\\ 
    39 &       30.73 &       20.85 &        9.87\\ 
    40 &       30.13 &       20.47 &        9.66\\ 
    41 &       27.71 &       20.32 &        7.39\\ 
    42 &       28.34 &       20.05 &        8.29\\ 
    43 &       27.27 &       19.84 &        7.43\\ 
    44 &       26.36 &       19.64 &        6.72\\ 
    45 &       25.15 &       19.42 &        5.73\\ 
    46 &       29.11 &       19.16 &        9.96\\ 
    47 &       25.38 &       18.99 &        6.38\\ 
    48 &       27.70 &       18.69 &        9.01\\ 
    49 &       24.93 &       18.59 &        6.33\\ 
    50 &       28.01 &       18.30 &        9.71\\ 
    51 &       24.30 &       18.20 &        6.10\\ 
    52 &       24.24 &       18.05 &        6.19\\ 
    53 &       22.56 &       17.88 &        4.68\\ 
    54 &       25.58 &       17.68 &        7.90\\ 
    55 &       24.06 &       17.52 &        6.54\\ 
    56 &       24.89 &       17.35 &        7.54\\ 
    57 &       23.98 &       17.18 &        6.80\\ 
    58 &       23.72 &       17.02 &        6.70\\ 
    59 &       23.69 &       16.90 &        6.79\\ 
    60 &       24.67 &       16.77 &        7.90\\ 
    61 &       20.50 &       16.64 &        3.86\\ 
    62 &       20.43 &       16.49 &        3.94\\ 
    63 &       21.60 &       16.37 &        5.23\\ 
    64 &       21.34 &       16.19 &        5.15\\ 
    65 &       20.78 &       16.11 &        4.66\\ 
    66 &       21.61 &       15.96 &        5.66\\ 
    67 &       21.32 &       15.86 &        5.46\\ 
    68 &       21.74 &       15.72 &        6.02\\ 
    69 &       20.41 &       15.60 &        4.82\\ 
    70 &       21.47 &       15.50 &        5.98\\ 
    71 &       21.46 &       15.39 &        6.07\\ 
    72 &       23.06 &       15.14 &        7.92\\ 
    73 &       17.51 &       15.12 &        2.39\\ 
    74 &       17.95 &       15.03 &        2.92\\ 
    75 &       20.02 &       14.95 &        5.07\\ 
    76 &       19.08 &       14.85 &        4.23\\ 
    77 &       22.03 &       14.74 &        7.29\\ 
    78 &       19.14 &       14.66 &        4.49\\ 
    79 &       19.77 &       14.56 &        5.21\\ 
    80 &       19.94 &       14.45 &        5.49\\ 
    81 &       19.07 &       14.38 &        4.69\\ 
    82 &       20.40 &       14.29 &        6.11\\ 
    83 &       17.08 &       14.22 &        2.86\\ 
    84 &       19.93 &       14.12 &        5.81\\ 
    85 &       19.92 &       14.05 &        5.88\\ 
    86 &       19.89 &       13.96 &        5.93\\ 
    87 &       18.70 &       13.88 &        4.81\\ 
    88 &       19.80 &       13.79 &        6.01\\ 
    89 &       19.76 &       13.71 &        6.05\\ 
    90 &       19.83 &       13.62 &        6.21\\ 
    91 &       17.18 &       13.56 &        3.62\\ 
    92 &       17.83 &       13.49 &        4.34\\ 
    93 &       18.75 &       13.43 &        5.32\\ 
    94 &       19.15 &       13.35 &        5.81\\ 
    95 &       18.54 &       13.29 &        5.25\\ 
    96 &       19.21 &       13.21 &        6.00\\ 
    97 &       18.38 &       13.14 &        5.24\\ 
    98 &       18.99 &       13.06 &        5.93\\ 
    99 &       19.06 &       13.00 &        6.06\\ 
   100 &       18.58 &       12.94 &        5.64\\ 
   101 &       18.67 &       12.87 &        5.80\\ 
   102 &       17.63 &       12.81 &        4.82\\ 
   103 &       17.51 &       12.74 &        4.77\\ 
   104 &       18.49 &       12.67 &        5.82\\ 
   105 &       18.94 &       12.62 &        6.32\\ 
   106 &       17.66 &       12.56 &        5.10\\ 
   107 &       18.62 &       12.50 &        6.13\\ 
   108 &       18.10 &       12.43 &        5.67\\ 
   109 &       18.23 &       12.38 &        5.85\\ 
   110 &       18.40 &       12.30 &        6.10\\ 
   111 &       18.70 &       12.25 &        6.45\\ 
   112 &       18.71 &       12.19 &        6.52\\ 
   113 &       18.28 &       12.15 &        6.14\\ 
   114 &       16.87 &       12.10 &        4.78\\ 
   115 &       17.42 &       12.05 &        5.37\\ 
   116 &       16.27 &       11.99 &        4.28\\ 
   117 &       17.67 &       11.94 &        5.73\\ 
   118 &       17.71 &       11.89 &        5.83\\ 
   119 &       17.09 &       11.84 &        5.25\\ 
   120 &       17.25 &       11.79 &        5.47\\ 
   121 &       17.56 &       11.73 &        5.83\\ 
   122 &       17.70 &       11.68 &        6.02\\ 
   123 &       14.38 &       11.64 &        2.74\\ 
   124 &       14.14 &       11.59 &        2.55\\ 
   125 &       17.38 &       11.54 &        5.84\\ 
   126 &       17.07 &       11.49 &        5.58\\ 
   127 &       16.54 &       11.45 &        5.09\\ 
   128 &       16.50 &       11.41 &        5.09\\ 
   129 &       16.98 &       11.36 &        5.62\\ 
   130 &       16.95 &       11.32 &        5.63\\ [1.0ex]
   \caption{Packing and covering radii for different cluster sizes
     $n$, measured in degrees, and the differences between them.
     Because the differences are always positive, we conclude that the
     optimal covering configuration is also the optimal minimal
     parking configuration, at least for $N \in \{ 4, \dots, 130
     \}$.}\label{tbl:coverings}
\end{longtable}


\end{document}